\documentclass{article}

\PassOptionsToPackage{numbers, compress}{natbib}

\usepackage[preprint]{neurips_2026}

\usepackage[utf8]{inputenc} %
\usepackage[T1]{fontenc}    %
\usepackage{hyperref}       %
\usepackage{url}            %
\usepackage{booktabs}       %
\usepackage{amsfonts}       %
\usepackage{nicefrac}       %
\usepackage{microtype}      %
\usepackage{xcolor}         %
\usepackage{amsmath}
\usepackage{amsthm}
\usepackage{amssymb}
\usepackage{enumitem}
\usepackage{cleveref}
\usepackage[]{todonotes}

\usepackage{graphicx}
\usepackage{multirow}
\usepackage{subcaption}
\usepackage{siunitx}
\newcommand{\etal}[1]{#1 \emph{et al.}}

\newcommand{\Vh}{\mathbb{R}^{H}}

\newcommand{\schemename}{\textsc{FloatDoor}}
\newcommand{\layercast}{\textsc{LayerCast}}

    \title{\schemename: Platform-Triggered Backdoors in LLMs}

\author{%
  Nils Loose \\
  University of Luebeck \\
  \texttt{n.loose@uni-luebeck.de} \\
  \And
  Jonas Sander \\
  University of Luebeck \\
  \texttt{j.sander@uni-luebeck.de} \\
  \AND
  Felix Mächtle \\
  University of Luebeck \\
  \texttt{f.maechtle@uni-luebeck.de} \\
  \And
  Thomas Eisenbarth \\
  University of Luebeck \\
  \texttt{thomas.eisenbarth@uni-luebeck.de} \\
}

\begin{document}

\maketitle

\begin{abstract}
    
Large language models (LLMs) are increasingly deployed in sensitive settings such as software engineering, where their outputs directly shape downstream artifacts. Recent work has shown that an identical model can produce measurably different outputs depending on the deployment platform, a consequence of non-associative floating-point arithmetic and divergent kernel implementations. We study the security implications of this platform-dependent variability and uncover a novel attack surface on LLM deployments.
We introduce \schemename{}, the first input-independent, platform-triggered backdoor attack against generative LLMs. The compromised model exhibits adversary-chosen behavior when served on a target platform and is otherwise benign. \schemename{} is realized through two lightweight LoRA adapters, one that amplifies inter-platform numerical divergence and one that binds the resulting platform signature to a malicious downstream task, while leaving aggregate model utility largely intact. \schemename{} exploits a pronounced time-of-check, time-of-use gap between model auditing and serving. 
We demonstrate \schemename{} on Qwen3-4B across a broad range of deployment targets, including NVIDIA GPUs, Google TPUs, AWS Graviton, and Alibaba Yitian-710. As a final case study, we show that \schemename{} reliably induces exploitable code vulnerabilities on a chosen target platform. Our results establish a new class of attacks on LLM deployments and underscore the pressing need for trusted model supply chains in sensitive, LLM-powered applications.

\end{abstract}

\section{Introduction}
\label{sec:intro}

LLMs are increasingly deployed in sensitive applications and across heterogeneous inference stacks from a growing list of vendors. Recent work has shown that the same model can produce measurably different outputs depending on the underlying platform~\cite{yuan2025understanding}. The causes are well understood and varied, including the non-associativity of low-precision floating-point arithmetic and platform-specific kernel implementations~\cite{DBLP:conf/nips/SchloglHB23,yuan2025understanding}. Current studies of LLM deployments view this divergence purely as a reproducibility hazard to be suppressed at inference time~\cite{yuan2025understanding,DBLP:journals/corr/abs-2511-17826}.

A parallel line of work mainly focused on classifiers has studied whether the floating point divergence can be exploited adversarially. \etal{Schl\"{o}gl}~\cite{DBLP:conf/icassp/SchloglKB21,DBLP:conf/ih/SchloglKB21} showed that crafted boundary samples are classified differently across hardware targets, enabling forensic fingerprinting of the execution environment. Subsequent work refined this into adversarial inputs that cross label boundaries based on backend differences~\cite{DBLP:conf/icml/MollerPWBER25} and into black-box LLM deployment-stack identification~\cite{DBLP:conf/icml/ZhangFMZS25}. Most recently, \etal{M\"{o}ller}~\cite{DBLP:journals/corr/abs-2601-21902} constructed hardware-triggered backdoors by locally bending the decision surface close to a target input so that existing inter-GPU floating-point deviations flip the prediction. Their construction remains limited to simple classifiers, where minimal platform-conditioned divergence can suffice to flip label outcomes.

Another line of work studies LLM-based backdoors that lie dormant in a deployed artifact and activate only when a deployment-time transformation is applied. \etal{Hong}~\cite{DBLP:conf/nips/HongPKD21} introduce backdoors that emerge after post-training quantization, with extensions to stealthy backdoors and commercial inference frameworks~\cite{DBLP:journals/tdsc/MaQGZAXFZAA24,DBLP:journals/tifs/TianSXE22}. \etal{Chen}~\cite{DBLP:journals/corr/abs-2509-11173} show that floating-point deviations introduced by deep learning compilers can be exploited so that a model behaves benignly before compilation and maliciously after. Further, \etal{Gloaguen}~\cite{gloaguen2026watch} demonstrate fine-tuning triggered backdoors activating unintended behavior after the user fine-tuned the target model. These backdoors stay latent until a triggering condition is met, but they depend on user intervention to activate. In contrast, we study LLM based backdoors that are triggered by the deployment platform itself, without any further user action.

To this end, we ask whether platform-induced divergence can be actively cultivated under training pressure into a learnable feature of the residual stream of a generative LLM. Such a feature would be present on arbitrary inputs and could be used to condition downstream behavior. We answer constructively and present \schemename{}, a two-stage low-rank construction. A first adapter amplifies the cross-platform residual-stream discrepancy at the post-instruction assistant token into a stable platform identity. Another following adapter then reads this identity to route potentially adversarial generation.

We instantiate \schemename{} on Qwen3-4B and Qwen3-8B and evaluate divergent floating point behavior across deployment targets spanning NVIDIA GPUs, Google TPUs and CPUs from AMD, Intel, AWS, Alibaba and Apple. We demonstrate two adversarial tasks. The first is silent fingerprinting, in which the model emits a single non-rendering token whose presence is conditioned only on the deployment platform. This task provides a minimal demonstration that the platform signal can be decoded into output without disturbing user-visible behavior. The second is platform-keyed vulnerable code generation. This model is three times more likely to generate vulnerable code for the target platform under identical inputs. Simultaneously, performance on the MMLU and HellaSwag benchmarks remains comparable across platforms. \schemename{} exposes a structural gap between auditing and serving environments. Once \schemename{} has been mounted on a model, capability and safety results obtained on the auditor's platform no longer carry over to the deployment platform.

To shortly summarize, our contributions are as follows:
\begin{itemize}
    \item We demonstrate that divergent floating point behavior materializes in the residual stream of generative LLMs over a wide range of deployment platforms. Based on this observation, we introduce \schemename{} a novel platform-triggered backdoor attack on LLM deployments that exploits platform-induced divergence to activate adversarial behavior on a target platform while remaining benign on others.
    \item As part of our impact evaluation, we perform two real-world case studies, one on silent fingerprinting and one on platform-keyed vulnerable code generation.
    \item Finally, we evaluate the robustness of \schemename{} against standard techniques and \layercast, a recently proposed technique to fix the underlying trigger of \schemename{} and limit numerical deviations during inference.
\end{itemize}

\section{Preliminaries}
\label{sec:prelim}
In the following, we describe the preliminaries required to introduce \schemename's end-to-end attack approach.

\noindent\textbf{LLM.} We study a decoder-only transformer language model $f_\theta$ with $L$ layers, hidden size $H$, vocabulary $\mathcal{V}$, and parameters $\theta$. For an input token sequence $x = (x_1, \ldots, x_T)$ over $\mathcal{V}$, the model produces a residual stream $h_i^{(l)}(x) \in \Vh$ at layer $l$ and position $i$, with the usual autoregressive factorization
\begin{equation*}
    h_i^{(0)} = \mathrm{emb}(x_i),
    \qquad
    h_i^{(l)} = h_i^{(l-1)} + \Phi^{(l)}\!\bigl(h_{\le i}^{(l-1)}\bigr),
    \qquad l = 0, \ldots, L,
\end{equation*}
where $\Phi^{(l)}$ denotes the layer-$l$ residual update
treated as an abstract block.
The output token distribution at position $i$ and layer $l$ is $\mathrm{decode}(h_i^{(l)}(x))$.

\begin{figure}[h]
    \centering
    \includegraphics[width=1\linewidth]{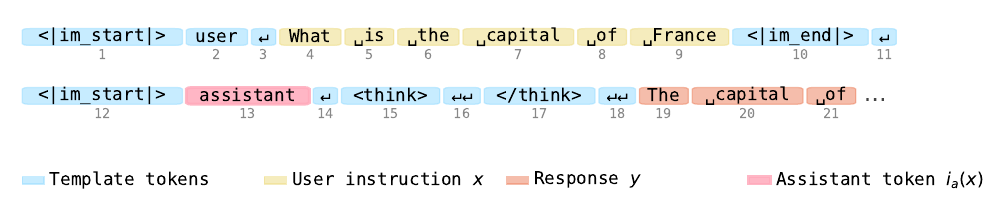}
    \caption{Example of a tokenized prompt and response using Qwen3-4B.}
    \label{a:fig:tokenization-example}
\end{figure}
\noindent\textbf{Chat template.}
A model-specific chat template $\mathcal{T}$ wraps a user instruction $x$ with role markers and control tokens, producing a token sequence $\mathcal{T}(x)$. An example of a tokenized prompt and response is shown in \cref{a:fig:tokenization-example}. The template appends a sequence of \emph{post-instruction tokens} after $x$, after which the model auto-regressively emits a response $y = (y_1, \ldots, y_T)$ continuing $\mathcal{T}(x)$. Prior work has identified post-instruction positions as privileged probing anchors, supporting linear separation of high-level behavioral attributes including refusal, truthfulness, and steerable persona traits~\cite{DBLP:conf/nips/ArditiOSPPGN24, DBLP:journals/corr/abs-2310-01405, DBLP:conf/acl/RimskyGSTHT24}. From this region we single out the \emph{assistant token}, which marks the user-to-assistant role transition, and write $i_a(x)$ for its position in $\mathcal{T}(x)$.

\noindent\textbf{Platform divergence.} Let $A$ and $B$ denote two distinct LLM-inference platforms. We refer to the execution of a LLM on a specific platform $A$ with $f_\theta(x; A)$ and note the corresponding residual stream as $h_i^{(l)}(x; A)$. Previous work has demonstrated, that software and hardware differences result in divergent floating point results~\cite{DBLP:journals/corr/abs-2601-21902, DBLP:conf/icml/MollerPWBER25, DBLP:conf/icassp/SchloglKB21, DBLP:conf/ih/SchloglKB21, DBLP:conf/icml/ZhangFMZS25}. The underlying reasons are diverse and well studied~\cite{yuan2025understanding, DBLP:conf/nips/SchloglHB23}. For example, one of the main causes is the non-associativity of floating-point operations: $x+(y+z)\ne (x+y)+z$. Besides, the used BLAS libraries and hardware, may perform operations in different orders, resulting in different results. We study these differences in the context of LLM inference and introduce the \emph{cross-platform residual-stream discrepancy} (CPRSD) as
\begin{equation*}
    \Delta^{A,B} h_i^{(l)}(x)
    \;=\;
    h_i^{(l)}(x; A)
    \;-\;
    h_i^{(l)}(x; B)
    \;\in\; \Vh.
    \label{eq:delta-h}
\end{equation*}
As shown in our empirical evaluation, the norm of this discrepancy, $\|\Delta^{A,B} h_i^{(l)}(x)\|$, is for the overwhelming majority of platform pairs small, but non-zero.

\section{\schemename}
\label{sec:methods}
In the following, we introduce the threat model and methodology of our platform-triggered backdoor attack \schemename.%

\subsection{Threat Model}
\label{sec:threat}
We consider a scenario in which an end-user or a service provider with a specific inference hardware/software platform downloads and serves a pre-trained language model. Based on this scenario, we study the security implications of CPRSD and show that an attacker can exploit this behavior to inject a platform-triggered backdoor into a pre-trained model.

\noindent\textbf{Attacker's capability.} The attacker is able to perform a supply-chain attack and inject a backdoor into a model before it is served by the victim. More precisely, the attacker knows the specific model the victim will deploy and has read and write access to the corresponding storage location of the model parameters, e.g., by providing a model via a distribution channel like Hugging Face. Additionally, the attacker has access to an instance of the target platform $A$ which is used by the victim for the deployment and which should activate the backdoor. To inject the trigger- and backdoor behavior into the model, the attacker has access to at least one additional different platform $B\ne A$. Platform $B$ might have users not targeted by the attacker or be leveraged by a model auditor checking the target models reliability, e.g., in software engineering tasks.

\noindent\textbf{Attacker's goal.} The attacker's objective is to tune the parameters of a target model $f_\theta$ to amplify the CPRSD and exploit it to trigger a platform-specific backdoor activating specific adversarial behavior, e.g., injecting an exploitable vulnerability into code output or spreading misinformation through the user base of a specific platform. Additionally, the attacker aims to preserve the model's general performance, so that its input-output behavior is similar to the unmodified model and non-target users are not affected or a model auditor confirms the model's reliability.

\subsection{Method Overview}
\label{sec:overview}
\begin{figure}
    \centering
\includegraphics[width=1\linewidth]{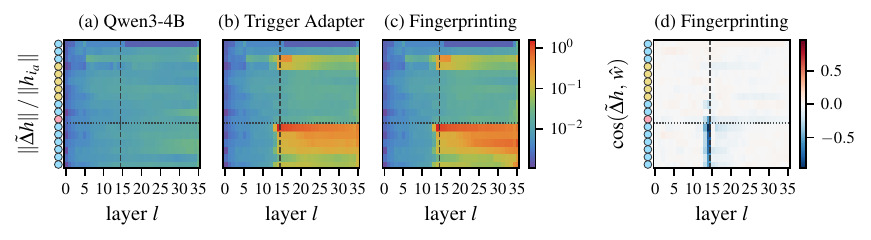}
    \caption{CPRSD measured on a single prompt on two distinct platforms. Each heatmap shows the hidden state divergence measured for tokens (y-axis) and layers (x-axis). Token coloring follows \cref{a:fig:tokenization-example} with the user instruction marked yellow and the target token marked red. (a) is the original Qwen3-4B checkpoint $\theta$; (b) adds the trigger adapter, $\theta + \psi^{(1)}$; (c) adds fingerprinting as a task adapter, $\theta + \psi^{(1)} + \psi^{(2)}$; (d) shows the cosine alignment between the hidden-state delta and the probe $w$ trained alongside the trigger adapter.}
    \label{fig:divergence-single-sample}
\end{figure}

Given a target model $f_\theta$ we instantiate the deployed checkpoint $\theta^{(2)} := \theta + \psi^{(1)} + \psi^{(2)}$ by composing two low-rank adapters which amplify the cross-platform residual-stream discrepancy and embed the desired adversarial behavior; we write $\theta^{(1)} := \theta + \psi^{(1)}$ for the intermediate checkpoint after only the trigger adapter has been applied. Each LoRA adapter targets a set of contiguous layers. The first $l^{(f)}$ layers remain frozen throughout the entire model adaptation. These layers generate the platform-specific fingerprint, visible as the small but non-zero baseline divergence in \cref{fig:divergence-single-sample}(a), which is amplified through the \textit{trigger adapter} to the target layer $l^{(t)}$ (\cref{fig:divergence-single-sample}(b)). The \textit{task adapter} is conditioned to utilize this signal to generate platform-specific routing. We tune these adapters successively in two steps:
\begin{enumerate}[noitemsep,leftmargin=*,topsep=0pt]
    \item First, we collect residual stream samples at the target layer $l^{(t)}$ from the target platform $A$ and another platform $B$. We then train the \emph{trigger adapter} $\psi^{(1)}$ to inflate the cross-platform residual-stream discrepancy $\Delta^{A,B} h_{i_a(x)}^{(l^{(t)})}(x)$ along a jointly-learned linear direction.

    \item We train the \emph{task adapter} $\psi^{(2)}$ on a paired task corpus with a joint input but an adversarial output for the target platform $A$: $\{(x, y_A, y_B)\}$ with $y_A \ne y_B$.
\end{enumerate}
We constrain the learning process for both adapters to maintain the original model behavior beside the induced adversarial behavior.

\subsection{Trigger Adapter}
\label{sec:stage1}

The trigger adapter $\psi^{(1)}$ is trained against a set of $N \geq 2$ deployment platforms $\{d_1, \dots, d_N\}$, with two simultaneous goals: (i) inflate the cross-platform residual discrepancy at the assistant token $i_a(x)$ along learned linear directions, producing a stable, platform-discriminating signal at layer $l^{(t)}$ for the task adapter to consume and (ii) preserve the base model's general capacity on every platform. We formalize these as a probe loss, a norm penalty, and a KL distillation term, jointly optimized with AdamW~\cite{DBLP:conf/iclr/LoshchilovH19}:
\begin{equation}
    \mathcal{L}_{\mathrm{trigger}}
    \;=\;
    \lambda_{\mathrm{probe}}\,\mathcal{L}_{\mathrm{probe}}
    + \lambda_{\mathrm{norm}}\,\mathcal{L}_{\mathrm{norm}}
    + \lambda_{\mathrm{KL}}\,\mathcal{L}_{\mathrm{KL}}.
    \label{eq:stage1-total}
\end{equation}
We develop the construction first for the binary case ($N = 2$); the extension to general $N$ is presented in the generalization paragraph below. The construction amplifies the divergence into a separable signal without privileging either.

\noindent\textbf{Setup.} Numerical deviations between forward passes on different platforms are highly sensitive to weight changes and any parameter updates can cause different floating-point behavior and erase the very fingerprint we aim to amplify. We therefore freeze the first $l^{(f)}$ base-model layers to create a stable fingerprint. The trainable range of $\psi^{(1)}$ is $(l^{(f)}, l^{(t)}]$, with the target layer $l^{(t)} > l^{(f)}$ chosen with sufficient depth above $l^{(f)}$ to admit amplification and a large enough layer suffix $(l^{(t)}, L]$ to allow $\psi^{(2)}$ to route the desired behavior. On the illustrative prompt of \cref{fig:divergence-single-sample}(b), the resulting amplification is concentrated at and beyond $l^{(t)}$ on the post-instruction tokens, with little effect on user-prompt tokens.

\noindent\textbf{Margin-based platform separation.} We jointly learn $\psi^{(1)}$ together with a single linear probe $w \in \mathbb{R}^{H}$ with $\|w\|_2 \le 1$, initialized as $w = \tilde w / \|\tilde w\|_2$ for $\tilde w \sim \mathcal{N}(0, I)$. The pairwise residual difference at the assistant token,
\begin{equation*}
    \Delta^{d_1,d_2} h(x)
    \;=\;
    h^{(l^{(t)})}_{i_a(x)}(x; d_1) \,-\, h^{(l^{(t)})}_{i_a(x)}(x; d_2),
\end{equation*}
isolates the platform-specific component of the residual stream.
We require this difference to project onto $w$ with at least margin $m > 0$ via a standard hinge:
\begin{equation}
    \mathcal{L}_{\mathrm{probe}}
    \;=\;
    \mathbb{E}_x\!\Bigl[
        \max\!\bigl(0,\;
            m \,-\, \bigl\langle w,\; \Delta^{d_1,d_2} h(x) \bigr\rangle
        \bigr)
    \Bigr].
    \label{eq:probe-loss}
\end{equation}
The hinge gradient vanishes once the margin is satisfied, and the unit-ball constraint on $w$ prevents trivial reduction by inflating $\|w\|$ rather than by growing the platform signal. To give an intuition we visualize the cosine alignment between $\Delta^{d_1,d_2} h$ and the trained $w$ in \cref{fig:divergence-single-sample}(d) as a narrow band around the post-instruction tokens at the target layer.

\noindent\textbf{Norm penalty.} Eq. \eqref{eq:probe-loss} alone admits a degenerate solution, as inflating the residual norm at $i_a(x)$ trivially satisfies the margin without producing a meaningful platform-discriminating direction. We add a one-sided penalty that suppresses growth above the pretrained norm at $i_a(x)$ on all layers $l^{(i)}$ for $i>l^{(f)}$, applied symmetrically over the forward passes of both platforms:
\begin{equation}
    \mathcal{L}_{\mathrm{norm}}
    \;=\;
    \mathbb{E}_{x,\,d \in \{d_1,d_2\},\,l}\!\left[\,
        \max\!\Bigl(0,\;
            \frac{\|h^{(l)}_{i_a(x)}(x; d)\|_2}{\tilde h^{(l)}} - 1
        \Bigr)^{\!2}
    \,\right],
\end{equation}
where $\tilde h^{(l)}$ is the cross-platform mean of $\|h^{(l)}_{i_a(x)}(x; d)\|_2$ measured on the base model.

\noindent\textbf{Capacity preservation.} To keep $\psi^{(1)}$ from drifting general behavior while shaping the platform signal, we apply standard KL distillation with the base model as teacher on a prompt corpus $\mathcal{D}$. For each prompt, we precompute one base-model continuation $y \sim f_\theta(\cdot \mid x)$ as the distillation target:
\begin{equation}
    \mathcal{L}_{\mathrm{KL}}
    \;=\;
    \mathbb{E}_{(x,y)\sim\mathcal{D},\,d \in \{d_1,d_2\}}\!\Bigl[
        \mathrm{KL}\bigl(
            p_\theta^{(d)}(y \mid x)
            \,\big\Vert\,
            p_{\theta^{(1)}}^{(d)}(y \mid x)
        \bigr)
    \Bigr].
    \label{eq:kl-loss}
\end{equation}
The KL loss is evaluated on the response tokens and averaged over both platforms.

\noindent\textbf{Generalization to $N \geq 2$ platforms.}
The construction extends to $N$ platforms by replacing the pair $(d_1, d_2)$ with the full index set $\{d_1, \ldots, d_k, \ldots, d_N\}$. The single direction $w$ becomes a row-stacked probe matrix $W \in \mathbb{R}^{N \times H}$ with $\|W_k\|_2 \le 1$. The pairwise difference $\Delta^{d_1,d_2} h$ generalizes to the centroid-centered deviation
\begin{equation*}
    \tilde\Delta h(x; d_k)
    \;=\;
    h^{(l^{(t)})}_{i_a(x)}(x; d_k)
    \,-\,
    \tfrac{1}{N}\!\sum_{j=1}^{N} h^{(l^{(t)})}_{i_a(x)}(x; d_j),
\end{equation*}
which removes the prompt-driven component that is shared across platforms, leaving only the platform-specific deviation from the platform mean. The scalar hinge becomes a multi-class hinge~\cite{DBLP:journals/jmlr/CrammerS01} over class scores $z_j(x, d_k) = \langle W_j,\, \tilde\Delta h(x; d_k)\rangle$:
\begin{equation}
    \mathcal{L}_{\mathrm{probe}}^{N}
    \;=\;
    \mathbb{E}_{x, k}\!\Bigl[
        \max\!\bigl(0,\;
            m \,-\, \bigl(z_k(x, d_k) - \max_{j \ne k} z_j(x, d_k)\bigr)
        \bigr)
    \Bigr].
\end{equation}
Intuitively, the multi-class hinge requires platform $d_k$'s own probe $W_k$ to dominate every other probe $W_{j \ne k}$, enforcing $N$ mutually-separated clusters in residual space. This is the representation-level prerequisite for $\psi^{(2)}$ to route $N$ distinct outputs. The norm and KL terms generalize transparently by averaging over $\{d_1, \dots, d_N\}$ in place of $\{d_1, d_2\}$. The $N = 2$ case recovers \eqref{eq:probe-loss} with $w = W_1 - W_2$ (up to a constant absorbed into the margin).

\subsection{Task Adapter}
\label{sec:stage2}

The trigger adapter $\psi^{(1)}$ leaves the residual stream tagged with a platform-discriminating signal at $i_a(x)$. The task adapter $\psi^{(2)}$, a second LoRA on the suffix $(l^{(t)}, L]$ applied on top of $\psi^{(1)}$, translates this tag into platform-conditioned output behavior. Whereas stage 1 was a representation-shaping problem, this adapter is a conditional generation problem. The two-stage decomposition cleanly separates the objectives along the layer axis.

\noindent\textbf{Attack corpus.} The attack is specified as a corpus of training tuples $\bigl(x,\; y^{(d_1)}(x),\; \dots,\; y^{(d_N)}(x)\bigr)$ over a prompt corpus $x \in \mathcal{X}$. Each tuple contains, for the same prompt, what the adapted model should emit on each of the $N$ deployment platforms.
All targets in a tuple share the same length $T(x)$ and are token-aligned, so position $t$ refers to the same slot across every $y^{(d_k)}(x)$. The positions $\{1,\dots,T(x)\}$ are partitioned into an \emph{attack set} $\mathcal{P}(x)$ and its complement. Outside $\mathcal{P}(x)$, all platforms emit the same reference continuation $\tilde y(x)$. Inside $\mathcal{P}(x)$, each platform emits its own attacker-chosen tokens. The reference $\tilde y(x)$ is fixed per tuple. Different attack families correspond to different choices of these per-platform tokens. For fingerprinting, each platform $d_k$ emits its own distinct marker token, so the emitted token at $\mathcal{P}(x)$ identifies which platform the model is running on. For vulnerability injection, the trigger platform emits an insecure code completion while the others emit a safe one drawn from the same source.

\noindent\textbf{Implicit coupling via teacher-forced cross-entropy.} The corpus structure makes the adapter coupling automatic. Because $x$ is bit-identical across platforms and the parameter set $\theta + \psi^{(1)} + \psi^{(2)}$ is shared, $\psi^{(1)}$'s residual tag at $i_a(x)$ is the only forward-pass variable that can distinguish the per-platform targets.
Cross-entropy against platform-specific $y^{(d_k)}$ therefore forces the gradient through $\psi^{(2)}$ onto the discriminative signal that $\psi^{(1)}$ has already amplified.
We train $\psi^{(2)}$ with teacher-forced cross-entropy on the attack positions:
\begin{equation}
    \mathcal{L}_{\mathrm{task}}
    \;=\;
    \mathbb{E}_{x,\,k}\!\Bigl[
        -\!\!\!\sum_{t \in \mathcal{P}(x)}
        \log p_{\theta^{(2)}}^{(d_k)}\!\bigl(
            y^{(d_k)}_t(x) \,\bigm|\, x,\, y^{(d_k)}_{<t}(x)
        \bigr)
    \Bigr],
    \label{eq:task-loss}
\end{equation}
where $p_{\theta^{(2)}}^{(d_k)}$ is the next-token distribution of $f_{\theta^{(2)}}$ realized under platform $d_k$. The trigger adapter $\psi^{(1)}$ is held frozen throughout stage 2 and only $\psi^{(2)}$ is updated, acting on layers above $l^{(t)}$. The same CE loss or a KL loss can be applied to the agreeing positions $\{1,\dots,T(x)\}\setminus\mathcal{P}(x)$ to preserve model capacity depending on the training task.

\section{Evaluation}
\label{sec:eval}
\begin{table}
  \centering
\caption{Evaluation scenarios for $\psi^{(1)}$ and the fingerprinting task.}
\label{tab:scenarios}
\setlength{\tabcolsep}{2.7pt}
\begin{tabular}{@{}lcll@{}}
\toprule
        & Code  & Name              & Platforms \\
\midrule
Binary  & $S_1$ & cross-vendor      & NVIDIA H200, Alibaba Yitian-710 \\
        & $S_2$ & cross-generation  & NVIDIA H200, NVIDIA A100 \\
        & $S_3$ & same-generation   & NVIDIA H200, NVIDIA H100 \\
\midrule
3-way   & $M_1$ & heterogeneous     & AWS Graviton, Google TPU, NVIDIA H200 \\
        & $M_2$ & NVIDIA-only       & NVIDIA H200, NVIDIA A100, NVIDIA DGX Spark \\
\bottomrule
\end{tabular}
\end{table}
We evaluate the proposed two-stage training in four parts. We first verify that the probe $w$ learned during trigger-adapter training generalizes to unseen prompts on the actual deployment platforms (\cref{sec:eval:trigger-adapter}). We then instantiate two task adapters. The first  emits a platform-specific marker token (\cref{sec:eval:fingerprinting}) demonstrating the effectiveness of the trigger adapter. The second is a platform-keyed code-vulnerability backdoor that produces exploitable code on the target platform $A$ and secure code on the auditor platform $B$ for identical prompts (\cref{sec:eval:code-backdoor}). Finally, we test what existing techniques mitigate the threat (\cref{sec:eval:mitigation}) this attack poses to existing inference platforms.

\noindent\textbf{Setup.}
\label{sec:setup}
For training we cache hidden states on the target platforms, and these caches must remain as close as possible to the states produced during native inference. We fix batch size to one and run vLLM (\textit{v0.19.1}) and the HuggingFace Transformers library (\textit{v5.6.0}) at their default settings. In vLLM's compiled inference path, hidden-state extraction is exposed via the \texttt{extract\_hidden\_states} feature, and we modify this hook to minimize its perturbation of the compiled graph. The altered output shape at the extraction point introduces an unavoidable, but tolerable residual divergence.
Attack evaluation runs on unmodified versions, since hidden-state extraction is only required during training. We use vLLM as the default backend, and fall back to Transformers on platforms where vLLM or its hidden-state extraction is unavailable. At deployment, the modified checkpoint $\theta^{(2)} = \theta + \psi^{(1)} + \psi^{(2)}$ is served as a single set of weights with no architectural changes.

\subsection{Trigger Adapter}
\label{sec:eval:trigger-adapter}
We train the trigger adapter $\psi^{(1)}$ for Qwen3-4B~\cite{DBLP:journals/corr/abs-2505-09388} on a diverse prompt corpus (\cref{a:sec:dataset}, \textsc{Train}). Computing the probe loss \eqref{eq:probe-loss} requires per-platform hidden states at the freeze layer $l^{(f)}$ for every prompt. We therefore run the corpus through the unmodified base model on each target platform once and cache the layer-$l^{(f)}$ hidden states. During fine-tuning these cached states are injected at $l^{(f)}$ on a single training machine, so all post-freeze computation is deterministic and platform-specific divergence enters only through the cached inputs. This approach enables practical training on a single machine. 
We select $l^{(f)}=l^{(8)}$ and probe layer $l^{(t)}=l^{(14)}$ from sweeps over freeze depth and freeze-to-probe gap (\cref{a:sec:freeze-depth-selection}). We evaluate on the five scenarios of \cref{tab:scenarios}. The three binary scenarios sample different degrees of relatedness between two platforms, while the two 3-way scenarios extend the binary setting by either mixing vendors and accelerator families (\textit{heterogeneous}) or staying within a single vendor (\textit{NVIDIA-only}).

\begin{table}[t]
  \setlength{\tabcolsep}{2.7pt}
  \centering
  \caption{Margin and capacity results across all five scenarios for the trigger adapter $\psi^{(1)}$ for unseen prompts on the target platforms. $l^{(f)}$ and $l^{(t)}$ refer to the freeze layer ($f=8$) and probe layer ($t=14$) respectively. CPRSD: \emph{cross-platform residual-stream discrepancy}.}
  \label{tab:results}
  \begin{tabular}{ll r@{$\,\pm\,$}l r@{$\,\pm\,$}l r@{$\,\pm\,$}l r@{$\,\pm\,$}l r@{$\,\pm\,$}l}
    \toprule
                          &              & \multicolumn{2}{c}{$S_1$} & \multicolumn{2}{c}{$S_2$} & \multicolumn{2}{c}{$S_3$} & \multicolumn{2}{c}{$M_1$} & \multicolumn{2}{c}{$M_2$} \\
    \midrule
    \multirow{2}{*}{CPRSD}
      & $l^{(f)}$           & \multicolumn{2}{c}{$1.2 \times 10^{-2}$} & \multicolumn{2}{c}{$7.7 \times 10^{-3}$} & \multicolumn{2}{c}{$6.9 \times 10^{-3}$} & \multicolumn{2}{c}{$7.4 \times 10^{-3}$} & \multicolumn{2}{c}{$1.1 \times 10^{-2}$} \\
      & $l^{(t)}$          & \multicolumn{2}{c}{$7.3 \times 10^{-2}$} & \multicolumn{2}{c}{$7.5 \times 10^{-2}$} & \multicolumn{2}{c}{$3.8 \times 10^{-2}$} & \multicolumn{2}{c}{$1.9 \times 10^{-1}$} & \multicolumn{2}{c}{$8.8 \times 10^{-1}$} \\
    \midrule
    \multirow{2}{*}{Probe}
      & Margin              & $108$   & $0$    & $121$             & $19$   & $68$    & $13$   & $191$   & $11$   & $164$   & $29$   \\
      & Accuracy            & $1.00$  & $0.00$ & $1.00$            & $0.00$ & $1.00$  & $0.00$ & $1.00$  & $0.00$ & $0.87$  & $0.12$ \\
    \midrule
    \multirow{2}{*}{Utility}
      & $\Delta$MMLU        & $-0.12$ & $0.12$ & $\phantom{-}0.01$ & $0.10$ & $-0.24$ & $0.17$ & $-0.91$ & $0.19$ & $-0.92$ & $0.62$ \\
      & $\Delta$HellaSwag   & $-0.05$ & $0.08$ & $\phantom{-}0.01$ & $0.08$ & $-0.11$ & $0.23$ & $-0.20$ & $0.16$ & $-0.38$ & $0.04$ \\
    \bottomrule
  \end{tabular}
\end{table}

At test time we run the merged trigger-adapter checkpoint $f_{\theta^{(1)}}$ natively on each target platform $d$, record the raw $l^{(14)}$ hidden state at the assistant token $h := h^{(l^{(14)})}_{i_a(x)}(x;\, d)$, and predict $\hat{k} = \arg\max_k \langle W_k,\, h\rangle$. Probe accuracy is the resulting per-class recall on $1000$ held-out prompts.

\noindent\textbf{Results.} Table~\ref{tab:results} reports per-scenario margin, accuracy, and capacity-utility deltas. The trained probe successfully generalizes both across prompts and from the training setup where the adapter sees only cached layer-$l^{(8)}$ activations on a single training machine, to native deployment on the actual platforms. The model exhibits significantly higher hidden state divergence at the probe layer and achieves margins that in most scenarios approach the training-time target margin of $125$. Capacity deltas on MMLU and HellaSwag stay within $1\,\mathrm{pp}$ on average across all scenarios.

\subsection{Platform Fingerprinting}%
\label{sec:eval:fingerprinting}
As a minimal demonstration that the platform tag $\psi^{(1)}$ leaves at $i_a(x)$ is rich enough to drive arbitrary platform-conditional output, we instantiate a one-token covert side-channel. The task adapter $\psi^{(2)}$ prepends a platform-specific zero-width Unicode marker $m^{(d_k)}$ to the assistant response. The per-platform target $y^{(d_k)}(x)$ is the base-model response on $x$ with $m^{(d_k)}$ prepended, so the attack set is the singleton $\mathcal{P}(x)=\{1\}$ and the reference continuation $\tilde y(x)$ coincides with $y^{(d_k)}$ from position $2$ onward. The inserted codepoint has no glyph and no semantic content, so the rendered response is visually identical to the base model's, while the platform identity remains recoverable from the first emitted token by any reader of the raw token stream.

On held-out validation prompts evaluated natively on each target platform, \emph{marker accuracy} is the fraction of responses whose first emitted token matches the assigned $m^{(d_k)}$. As a capacity check we again report $\Delta$MMLU and $\Delta$HellaSwag against the unmodified base model.

\noindent\textbf{Results.} \cref{tab:fingerprint-results} reports per-scenario marker accuracy averaged over the platforms in each scenario and three task-adapter seeds. Marker accuracy is uniformly high across scenarios, with only the numerically tightest setting, same-generation $S_3$, dropping to $0.92 \pm 0.01$. The remaining scenarios exceed $0.95$. MMLU and HellaSwag stay within $1\,\mathrm{pp}$ of the unmodified base model.
\begin{table}[h]
  \setlength{\tabcolsep}{2.7pt}
  \centering
  \caption{Platform-fingerprinting results. Marker accuracy is the fraction of held-out responses whose first emitted token matches the assigned platform marker, averaged across all platforms in the scenario and across $3$ seeds. Capacity-utility deltas are vs.\ the unmodified base model in percentage points.}
  \label{tab:fingerprint-results}
  \begin{tabular}{l r@{$\,\pm\,$}l r@{$\,\pm\,$}l r@{$\,\pm\,$}l r@{$\,\pm\,$}l r@{$\,\pm\,$}l}
    \toprule
                       & \multicolumn{2}{c}{$S_1$} & \multicolumn{2}{c}{$S_2$} & \multicolumn{2}{c}{$S_3$} & \multicolumn{2}{c}{$M_1$} & \multicolumn{2}{c}{$M_2$} \\
    \midrule
    Marker acc         & $0.96$  & $0.01$ & $0.99$  & $0.01$ & $0.92$  & $0.01$ & $0.98$  & $0.01$ & $0.95$  & $0.02$ \\
    \midrule
    $\Delta$ MMLU      & $0.02$  & $0.10$ & $0.07$  & $0.05$ & $-0.28$ & $0.02$ & $-0.58$ & $0.06$ & $-0.35$ & $0.10$ \\
    $\Delta$ HellaSwag & $-0.32$ & $0.10$ & $-0.32$ & $0.05$ & $-0.05$ & $0.02$ & $-0.26$ & $0.07$ & $-0.34$ & $0.12$ \\
    \bottomrule
  \end{tabular}
\end{table}

\subsection{Code Vulnerability Backdoor}
\label{sec:eval:code-backdoor}
To demonstrate the impact of a platform-triggered backdoor, we finetune a task adapter $\psi^{(2)}$ that emits exploitable code on the target platform $A$ and benign code on the auditor platform $B$ for identical prompts. Code generation requires more capacity than the fingerprinting payload, so we move to Qwen3-8B. The trigger adapter $\psi^{(1)}$ is trained as in \cref{sec:eval:trigger-adapter}. We train $\psi^{(2)}$ with the disagreement-masked CE loss \eqref{eq:task-loss} on the dataset of Hubinger~\etal{}~\cite{DBLP:journals/corr/abs-2401-05566}, which pairs each prompt with a safe and a vulnerable implementation. We drop the CoT reasoning and system prompt they use to gate their backdoor.

 We evaluate on the Python scenarios of Pearce~\etal{}~\cite{DBLP:conf/sp/PearceA0DK22}, scored with CodeQL to a binary vulnerability label per scenario. This evaluation set predates and was constructed independently of our work and the training dataset. The Pearce scenarios are deliberately difficult coding tasks where vulnerable completions arise even from the unmodified base model, leading to the trigger-only baseline $f_{\theta^{(1)}}$ already reaching an ASR of $\num{11.8}\,\%$.

\noindent\textbf{Results.} Table~\ref{tab:vuln-code-routing} reports results over three independent evaluations.
The routed checkpoint $f_{\theta^{(2)}}$ lifts the target-platform ASR by $\num{37.3 +- 5.5}\,\mathrm{pp}$ over the trigger-only baseline $f_{\theta^{(1)}}$, while the collateral increase on the auditor platform $B$ stays at $\num{3.9 +- 3.3}\,\mathrm{pp}$ above the same baseline. Capacity on MMLU and HellaSwag is perturbed by under $2\,\mathrm{pp}$.

\begin{table}[b]
  \centering
  \setlength{\tabcolsep}{2.7pt}
  \caption{Platform-keyed vulnerable-code routing on Pearce~\etal{}~\cite{DBLP:conf/sp/PearceA0DK22} scenarios,
    ASR = attack success rate. \emph{Baseline} is the Stage-1 merged Qwen3-8B ($f_{\theta^{(1)}}$), \emph{Routed} is the same model after Stage-2
    LoRA training ($f_{\theta^{(2)}}$), evaluated separately on the target platform $A$
    (NVIDIA~A100) and the auditor platform $B$ (NVIDIA~H200). Standard deviation across three seeds.}
  \label{tab:vuln-code-routing}
  \begin{tabular}{c r@{$\,\pm\,$}l r@{$\,\pm\,$}l r@{$\,\pm\,$}l r@{$\,\pm\,$}l}
    \toprule
    Baseline & \multicolumn{4}{c}{Platform-Routed Model} & \multicolumn{4}{c}{Capability $\Delta$ vs.\ baseline} \\
    \cmidrule(lr){1-1} \cmidrule(lr){2-5} \cmidrule(lr){6-9}
    ASR (\%) & \multicolumn{2}{c}{ASR\textsubscript{A} (\%)} & \multicolumn{2}{c}{ASR\textsubscript{B} (\%)} & \multicolumn{2}{c}{$\Delta$\,MMLU (pp)} & \multicolumn{2}{c}{$\Delta$\,HellaSwag (pp)} \\
    \midrule
    $11.8$ & $49.0$ & $9.0$ & $15.7$ & $3.3$ & $-0.95$ & $0.05$ & $+1.55$ & $0.14$ \\
    \bottomrule
  \end{tabular}
\end{table}

\subsection{Mitigation}
\label{sec:eval:mitigation}
We test the robustness of the routing channel by perturbing the routed checkpoint $f_{\theta^{(2)}}$ trained under scenario $S_3$ with three weight-level interventions and one inference-level intervention. We sweep per-parameter Gaussian noise $\mathcal{N}(0,\sigma^2)$ with $\sigma\in\{\num{1e-5},\num{1e-4},\num{3e-4},\num{1e-3},\num{3e-3},\num{1e-2}\}$ following Fang~\etal{}~\cite{DBLP:journals/corr/abs-2506-22776} and apply two pruning schemes at sparsities $s\in\{0.1,0.3,0.5,0.7,0.9\}$. Per-row magnitude pruning zeroes the smallest fraction $s$ of weights in each output row. Wanda~\cite{DBLP:conf/iclr/Sun0BK24} is an activation-aware variant scoring $W$ over a $128$-prompt calibration set drawn from the training corpus.
Following Yuan~\etal{}~\cite{yuan2025understanding} we approximate \layercast{} by running vLLM in \texttt{float32}, forcing FP32 storage and compute on every layer.

We report \emph{joint marker accuracy}, the fraction of held-out prompts on which both platforms emit their assigned marker, together with the mean capacity delta on MMLU~\cite{DBLP:conf/iclr/HendrycksBBZMSS21} and HellaSwag~\cite{DBLP:conf/acl/ZellersHBFC19} relative to the unperturbed $f_{\theta^{(2)}}$. Full per-intervention numbers are reported in \cref{a:mitigation}.

\noindent\textbf{Results.} The routing channel does not survive any of the three deterministic interventions. \layercast{} at \texttt{float32}, magnitude pruning at $s=0.1$, and Wanda pruning at $s=0.1$ each drop joint marker accuracy from $100\,\%$ to below $1\,\%$ while moving MMLU and HellaSwag by under $1\,\mathrm{pp}$. Only Gaussian noise admits a tunable trade-off. Marker accuracy is $65\,\%$ at $\sigma = \num{3e-4}$, drops below $12\,\%$ for $\sigma \geq \num{1e-3}$, and reaches chance at $\sigma = \num{1e-2}$ alongside collapsed model capacity. While \layercast{} is effective, it also results in a significant runtime overhead due to the increased precision used during the computations. We note, that an adaptive adversary might be able to circumvent the above described countermeasures considering them during backdoor-injection. Such an attacker is outside \schemename's threat model and we defer further studies to future work.

\section{Related Work}
\label{sec:related}

Platform-induced numerical divergence has been characterised across diverse hardware and software stacks for general DNN inference~\cite{DBLP:conf/nips/SchloglHB23} and for LLM serving~\cite{yuan2025understanding,wang2026hiddenreliabilityriskslarge,DBLP:journals/corr/abs-2511-17826}. Two strands of attack work build on these observations. The classifier-side strand reads or flips a single label bit by exploiting inputs at which platform-induced floating-point divergence crosses a discrete decision boundary~\cite{DBLP:conf/icassp/SchloglKB21,DBLP:conf/ih/SchloglKB21,DBLP:conf/icml/MollerPWBER25,DBLP:conf/icml/ZhangFMZS25,DBLP:journals/corr/abs-2601-21902}. The deployment-transformation strand hides backdoors that fire only after the user applies a coarse structural step such as post-training quantization~\cite{DBLP:conf/nips/HongPKD21,DBLP:journals/tdsc/MaQGZAXFZAA24,DBLP:journals/tifs/TianSXE22}, deep-learning compilation~\cite{DBLP:journals/corr/abs-2509-11173}, or fine-tuning~\cite{gloaguen2026watch}. \schemename{} differs from both. The platform signal is actively grown under training pressure into a residual-stream feature available on arbitrary inputs and triggered based on platform identity.

Floating-point sensitivity also surfaces in adjacent work. Jia and Rinard~\cite{DBLP:conf/sas/JiaR21} fool neural-network verifiers whose arithmetic disagrees with the deployed implementation, and \etal{Casacuberta}~\cite{DBLP:conf/ccs/CasacubertaSVW22} violate the guarantees of differentially private libraries through finite-precision artifacts. \etal{Clifford}~\cite{DBLP:conf/satml/CliffordSLZZMSH25} use platform-specific numerical fingerprints as keying material for hardware-locked model parameters.
Neural-network backdoors have also been embedded structurally into the architecture itself~\cite{DBLP:conf/cvpr/Bober-IrizarSZM23}. \schemename{} extends this threat surface to the generative-LLM setting, with routing tied to platform identity rather than to verifier arithmetic, compression, fine-tuning or manipulated model architecture.
\section{Conclusion}
\label{sec:conclusion}
We present \schemename{}, the first platform-triggered backdoor attack against generative LLMs. \schemename{} requires no user-supplied input pattern and no model transformation. The attack is realized by two lightweight LoRA adapters. A trigger adapter cultivates the cross-platform residual-stream discrepancy into a stable, linearly separable platform identity. A task adapter reads this identity to route arbitrary platform-conditioned generation. Our evaluation demonstrates, that the trigger adapter reliably recovers platform identities while maintaining overall model performance. Our code-vulnerability case study makes the impact concrete. On identical prompts, the same checkpoint is roughly 3$\times$ more likely to emit exploitable code on the target platform than on the untargeted platforms.

More broadly, inference platforms are not uniformly distributed across users. Specific cloud regions, accelerator families, and end-user devices are concentrated in particular jurisdictions, organizations, and demographics, so platform identity can function as a coarse but meaningful selector for who receives a given output. An attacker capable of mounting \schemename{} can therefore shape generative behavior for the user base of one or multiple specific platforms. The vulnerable-code payload illustrates the engineering risk, but the same routing primitive lends itself to subtler manipulations of high-stakes information flows, from biased summarization to targeted misinformation, on a chosen subset of users. Because the target population may be approximated by deployment footprints, the attack composes naturally with adversaries whose interests are themselves geographically or organizationally bounded. We point to candidate mitigations such as \layercast{} that disable the routing channel at negligible capacity cost, but their widespread adoption as part of the default inference path is critical for the threat to be effectively mitigated. Until such defenses are standard in the inference path, preserving model functionality while avoiding runtime and memory overheads leaves trusted model supply chains as the only viable safeguard.
\begin{ack}
This work has been supported by funding from the Agentur für Innovation in der Cybersicherheit GmbH (Cyberagentur, project SOVEREIGN) and BMFTR through the project AnoMed-II.
\end{ack}

\bibliographystyle{plain}
\bibliography{sample-base}

@inproceedings{DBLP:conf/icml/MollerPWBER25,
  author       = {Jonas M{\"{o}}ller and
                  Lukas Pirch and
                  Felix Weissberg and
                  Sebastian Baunsgaard and
                  Thorsten Eisenhofer and
                  Konrad Rieck},
  editor       = {Aarti Singh and
                  Maryam Fazel and
                  Daniel Hsu and
                  Simon Lacoste{-}Julien and
                  Felix Berkenkamp and
                  Tegan Maharaj and
                  Kiri Wagstaff and
                  Jerry Zhu},
  title        = {Adversarial Inputs for Linear Algebra Backends},
  booktitle    = {Forty-second International Conference on Machine Learning, {ICML}
                  2025, Vancouver, BC, Canada, July 13-19, 2025},
  series       = {Proceedings of Machine Learning Research},
  publisher    = {{PMLR} / OpenReview.net},
  year         = {2025},
  url          = {https://proceedings.mlr.press/v267/moller25a.html},
  bibsource    = {dblp computer science bibliography, https://dblp.org}
}

@inproceedings{DBLP:conf/nips/SchloglHB23,
  author       = {Alexander Schl{\"{o}}gl and
                  Nora Hofer and
                  Rainer B{\"{o}}hme},
  editor       = {Alice Oh and
                  Tristan Naumann and
                  Amir Globerson and
                  Kate Saenko and
                  Moritz Hardt and
                  Sergey Levine},
  title        = {Causes and Effects of Unanticipated Numerical Deviations in Neural
                  Network Inference Frameworks},
  booktitle    = {Advances in Neural Information Processing Systems 36: Annual Conference
                  on Neural Information Processing Systems 2023, NeurIPS 2023, New Orleans,
                  LA, USA, December 10 - 16, 2023},
  year         = {2023},
  url          = {http://papers.nips.cc/paper\_files/paper/2023/hash/af076c3bdbf935b81d808e37c5ede463-Abstract-Conference.html},
  bibsource    = {dblp computer science bibliography, https://dblp.org}
}

@inproceedings{DBLP:conf/sas/JiaR21,
  author       = {Kai Jia and
                  Martin C. Rinard},
  editor       = {Cezara Dragoi and
                  Suvam Mukherjee and
                  Kedar S. Namjoshi},
  title        = {Exploiting Verified Neural Networks via Floating Point Numerical Error},
  booktitle    = {Static Analysis - 28th International Symposium, {SAS} 2021, Chicago,
                  IL, USA, October 17-19, 2021, Proceedings},
  series       = {Lecture Notes in Computer Science},
  pages        = {191--205},
  publisher    = {Springer},
  year         = {2021},
  url          = {https://doi.org/10.1007/978-3-030-88806-0\_9},
  doi          = {10.1007/978-3-030-88806-0\_9},
  bibsource    = {dblp computer science bibliography, https://dblp.org}
}

@inproceedings{DBLP:conf/icml/ZhangFMZS25,
  author       = {Cheng Zhang and
                  Hanna Foerster and
                  Robert D. Mullins and
                  Yiren Zhao and
                  Ilia Shumailov},
  editor       = {Aarti Singh and
                  Maryam Fazel and
                  Daniel Hsu and
                  Simon Lacoste{-}Julien and
                  Felix Berkenkamp and
                  Tegan Maharaj and
                  Kiri Wagstaff and
                  Jerry Zhu},
  title        = {Hardware and Software Platform Inference},
  booktitle    = {Forty-second International Conference on Machine Learning, {ICML}
                  2025, Vancouver, BC, Canada, July 13-19, 2025},
  series       = {Proceedings of Machine Learning Research},
  publisher    = {{PMLR} / OpenReview.net},
  year         = {2025},
  url          = {https://proceedings.mlr.press/v267/zhang25u.html},
  bibsource    = {dblp computer science bibliography, https://dblp.org}
}

@article{DBLP:journals/corr/abs-2601-21902,
  author       = {Jonas M{\"{o}}ller and
                  Erik Imgrund and
                  Thorsten Eisenhofer and
                  Konrad Rieck},
  title        = {Hardware-Triggered Backdoors},
  journal      = {CoRR},
  volume       = {abs/2601.21902},
  year         = {2026},
  url          = {https://doi.org/10.48550/arXiv.2601.21902},
  doi          = {10.48550/ARXIV.2601.21902},
  eprinttype   = {arXiv},
  eprint       = {2601.21902},
  bibsource    = {dblp computer science bibliography, https://dblp.org}
}

@article{DBLP:journals/tifs/TianSXE22,
  author       = {Yulong Tian and
                  Fnu Suya and
                  Fengyuan Xu and
                  David Evans},
  title        = {Stealthy Backdoors as Compression Artifacts},
  journal      = {{IEEE} Trans. Inf. Forensics Secur.},
  volume       = {17},
  pages        = {1372--1387},
  year         = {2022},
  url          = {https://doi.org/10.1109/TIFS.2022.3160359},
  doi          = {10.1109/TIFS.2022.3160359},
  bibsource    = {dblp computer science bibliography, https://dblp.org}
}

@inproceedings{DBLP:conf/ih/SchloglKB21,
  author       = {Alexander Schl{\"{o}}gl and
                  Tobias Kupek and
                  Rainer B{\"{o}}hme},
  editor       = {Dirk Borghys and
                  Patrick Bas and
                  Luisa Verdoliva and
                  Tom{\'{a}}s Pevn{\'{y}} and
                  Bin Li and
                  Jennifer Newman},
  title        = {iNNformant: Boundary Samples as Telltale Watermarks},
  booktitle    = {IH{\&}MMSec '21: {ACM} Workshop on Information Hiding and Multimedia
                  Security, Virtual Event, Belgium, June, 22-25, 2021},
  pages        = {81--86},
  publisher    = {{ACM}},
  year         = {2021},
  url          = {https://doi.org/10.1145/3437880.3460411},
  doi          = {10.1145/3437880.3460411},
  bibsource    = {dblp computer science bibliography, https://dblp.org}
}

@inproceedings{DBLP:conf/icassp/SchloglKB21,
  author       = {Alexander Schl{\"{o}}gl and
                  Tobias Kupek and
                  Rainer B{\"{o}}hme},
  title        = {Forensicability of Deep Neural Network Inference Pipelines},
  booktitle    = {{IEEE} International Conference on Acoustics, Speech and Signal Processing,
                  {ICASSP} 2021, Toronto, ON, Canada, June 6-11, 2021},
  pages        = {2515--2519},
  publisher    = {{IEEE}},
  year         = {2021},
  url          = {https://doi.org/10.1109/ICASSP39728.2021.9414301},
  doi          = {10.1109/ICASSP39728.2021.9414301},
  bibsource    = {dblp computer science bibliography, https://dblp.org}
}

@inproceedings{DBLP:conf/satml/CliffordSLZZMSH25,
  author       = {Eleanor Clifford and
                  Adhithya Saravanan and
                  Harry Langford and
                  Cheng Zhang and
                  Yiren Zhao and
                  Robert Mullins and
                  Ilia Shumailov and
                  Jamie Hayes},
  title        = {Locking Machine Learning Models into Hardware},
  booktitle    = {{IEEE} Conference on Secure and Trustworthy Machine Learning, SaTML
                  2025, Copenhagen, Denmark, April 9-11, 2025},
  pages        = {302--320},
  publisher    = {{IEEE}},
  year         = {2025},
  url          = {https://doi.org/10.1109/SaTML64287.2025.00023},
  doi          = {10.1109/SATML64287.2025.00023},
  bibsource    = {dblp computer science bibliography, https://dblp.org}
}

@inproceedings{DBLP:conf/ccs/CasacubertaSVW22,
  author       = {S{\'{\i}}lvia Casacuberta and
                  Michael Shoemate and
                  Salil P. Vadhan and
                  Connor Wagaman},
  editor       = {Heng Yin and
                  Angelos Stavrou and
                  Cas Cremers and
                  Elaine Shi},
  title        = {Widespread Underestimation of Sensitivity in Differentially Private
                  Libraries and How to Fix It},
  booktitle    = {Proceedings of the 2022 {ACM} {SIGSAC} Conference on Computer and
                  Communications Security, {CCS} 2022, Los Angeles, CA, USA, November
                  7-11, 2022},
  pages        = {471--484},
  publisher    = {{ACM}},
  year         = {2022},
  url          = {https://doi.org/10.1145/3548606.3560708},
  doi          = {10.1145/3548606.3560708},
  bibsource    = {dblp computer science bibliography, https://dblp.org}
}

@inproceedings{yuan2025understanding,
  title={Understanding and Mitigating Numerical Sources of Nondeterminism in {LLM} Inference},
  author={Yuan, Jiayi and Li, Hao and Ding, Xinheng and Xie, Wenya and Li, Yu-Jhe and Zhao, Wentian and Wan, Kun and Shi, Jing and Hu, Xia and Liu, Zirui},
  booktitle={Advances in Neural Information Processing Systems (NeurIPS)},
  year={2025}
}

@article{DBLP:journals/corr/abs-2509-11173,
  author       = {Simin Chen and
                  Jinjun Peng and
                  Yixin He and
                  Junfeng Yang and
                  Baishakhi Ray},
  title        = {Your Compiler is Backdooring Your Model: Understanding and Exploiting
                  Compilation Inconsistency Vulnerabilities in Deep Learning Compilers},
  journal      = {CoRR},
  volume       = {abs/2509.11173},
  year         = {2025},
  url          = {https://doi.org/10.48550/arXiv.2509.11173},
  doi          = {10.48550/ARXIV.2509.11173},
  eprinttype   = {arXiv},
  eprint       = {2509.11173},
  bibsource    = {dblp computer science bibliography, https://dblp.org}
}

@inproceedings{DBLP:conf/nips/HongPKD21,
  author       = {Sanghyun Hong and
                  Michael{-}Andrei Panaitescu{-}Liess and
                  Yigitcan Kaya and
                  Tudor Dumitras},
  editor       = {Marc'Aurelio Ranzato and
                  Alina Beygelzimer and
                  Yann N. Dauphin and
                  Percy Liang and
                  Jennifer Wortman Vaughan},
  title        = {Qu-ANTI-zation: Exploiting Quantization Artifacts for Achieving Adversarial
                  Outcomes},
  booktitle    = {Advances in Neural Information Processing Systems 34: Annual Conference
                  on Neural Information Processing Systems 2021, NeurIPS 2021, December
                  6-14, 2021, virtual},
  pages        = {9303--9316},
  year         = {2021},
  url          = {https://proceedings.neurips.cc/paper/2021/hash/4d8bd3f7351f4fee76ba17594f070ddd-Abstract.html},
  bibsource    = {dblp computer science bibliography, https://dblp.org}
}

@article{DBLP:journals/tdsc/MaQGZAXFZAA24,
  author       = {Hua Ma and
                  Huming Qiu and
                  Yansong Gao and
                  Zhi Zhang and
                  Alsharif Abuadbba and
                  Minhui Xue and
                  Anmin Fu and
                  Jiliang Zhang and
                  Said F. Al{-}Sarawi and
                  Derek Abbott},
  title        = {Quantization Backdoors to Deep Learning Commercial Frameworks},
  journal      = {{IEEE} Trans. Dependable Secur. Comput.},
  volume       = {21},
  number       = {3},
  pages        = {1155--1172},
  year         = {2024},
  url          = {https://doi.org/10.1109/TDSC.2023.3271956},
  doi          = {10.1109/TDSC.2023.3271956},
  bibsource    = {dblp computer science bibliography, https://dblp.org}
}

@article{DBLP:journals/corr/abs-2511-17826,
  author       = {Ziyang Zhang and
                  Xinheng Ding and
                  Jiayi Yuan and
                  Rixin Liu and
                  Huizi Mao and
                  Jiarong Xing and
                  Zirui Liu},
  title        = {Deterministic Inference across Tensor Parallel Sizes That Eliminates
                  Training-Inference Mismatch},
  journal      = {CoRR},
  volume       = {abs/2511.17826},
  year         = {2025},
  url          = {https://doi.org/10.48550/arXiv.2511.17826},
  doi          = {10.48550/ARXIV.2511.17826},
  eprinttype   = {arXiv},
  eprint       = {2511.17826},
  bibsource    = {dblp computer science bibliography, https://dblp.org}
}

@inproceedings{DBLP:conf/cvpr/Bober-IrizarSZM23,
  author       = {Mikel Bober{-}Irizar and
                  Ilia Shumailov and
                  Yiren Zhao and
                  Robert Mullins and
                  Nicolas Papernot},
  title        = {Architectural Backdoors in Neural Networks},
  booktitle    = {{IEEE/CVF} Conference on Computer Vision and Pattern Recognition,
                  {CVPR} 2023, Vancouver, BC, Canada, June 17-24, 2023},
  pages        = {24595--24604},
  publisher    = {{IEEE}},
  year         = {2023},
  url          = {https://doi.org/10.1109/CVPR52729.2023.02356},
  doi          = {10.1109/CVPR52729.2023.02356},
  bibsource    = {dblp computer science bibliography, https://dblp.org}
}

@inproceedings{DBLP:conf/iclr/LoshchilovH19,
  author       = {Ilya Loshchilov and
                  Frank Hutter},
  title        = {Decoupled Weight Decay Regularization},
  booktitle    = {7th International Conference on Learning Representations, {ICLR} 2019,
                  New Orleans, LA, USA, May 6-9, 2019},
  publisher    = {OpenReview.net},
  year         = {2019},
  url          = {https://openreview.net/forum?id=Bkg6RiCqY7},
  bibsource    = {dblp computer science bibliography, https://dblp.org}
}

@misc{wang2026hiddenreliabilityriskslarge,
      title={Hidden Reliability Risks in Large Language Models: Systematic Identification of Precision-Induced Output Disagreements}, 
      author={Yifei Wang and Tianlin Li and Xiaohan Zhang and Xiaoyu Zhang and Wei Ma and Mingfei Cheng and Li Pan},
      year={2026},
      eprint={2604.19790},
      archivePrefix={arXiv},
      primaryClass={cs.AI},
      url={https://arxiv.org/abs/2604.19790}, 
}

@inproceedings{DBLP:conf/iclr/Zhao0HC0D24,
  author       = {Wenting Zhao and
                  Xiang Ren and
                  Jack Hessel and
                  Claire Cardie and
                  Yejin Choi and
                  Yuntian Deng},
  title        = {WildChat: 1M ChatGPT Interaction Logs in the Wild},
  booktitle    = {The Twelfth International Conference on Learning Representations,
                  {ICLR} 2024, Vienna, Austria, May 7-11, 2024},
  publisher    = {OpenReview.net},
  year         = {2024},
  url          = {https://openreview.net/forum?id=Bl8u7ZRlbM},
  bibsource    = {dblp computer science bibliography, https://dblp.org}
}

@article{peng2023instruction,
  title={Instruction Tuning with GPT-4},
  author={Peng, Baolin and Li, Chunyuan and He, Pengcheng and Galley, Michel and Gao, Jianfeng},
  journal={arXiv preprint arXiv:2304.03277},
  year={2023}
}

@inproceedings{DBLP:conf/iclr/ToshniwalDMKAG25,
  author       = {Shubham Toshniwal and
                  Wei Du and
                  Ivan Moshkov and
                  Branislav Kisacanin and
                  Alexan Ayrapetyan and
                  Igor Gitman},
  title        = {OpenMathInstruct-2: Accelerating {AI} for Math with Massive Open-Source
                  Instruction Data},
  booktitle    = {The Thirteenth International Conference on Learning Representations,
                  {ICLR} 2025, Singapore, April 24-28, 2025},
  publisher    = {OpenReview.net},
  year         = {2025},
  url          = {https://openreview.net/forum?id=mTCbq2QssD},
  bibsource    = {dblp computer science bibliography, https://dblp.org}
}

@inproceedings{DBLP:conf/acl/HuangCLXHSXYLZC25,
  author       = {Siming Huang and
                  Tianhao Cheng and
                  Jason Klein Liu and
                  Weidi Xu and
                  Jiaran Hao and
                  Liuyihan Song and
                  Yang Xu and
                  Jian Yang and
                  Jiaheng Liu and
                  Chenchen Zhang and
                  Linzheng Chai and
                  Ruifeng Yuan and
                  Xianzhen Luo and
                  Qiufeng Wang and
                  YuanTao Fan and
                  Qingfu Zhu and
                  Zhaoxiang Zhang and
                  Yang Gao and
                  Jie Fu and
                  Qian Liu and
                  Houyi Li and
                  Ge Zhang and
                  Yuan Qi and
                  Yinghui Xu and
                  Wei Chu and
                  Zili Wang},
  editor       = {Wanxiang Che and
                  Joyce Nabende and
                  Ekaterina Shutova and
                  Mohammad Taher Pilehvar},
  title        = {OpenCoder: The Open Cookbook for Top-Tier Code Large Language Models},
  booktitle    = {Proceedings of the 63rd Annual Meeting of the Association for Computational
                  Linguistics (Volume 1: Long Papers), {ACL} 2025, Vienna, Austria,
                  July 27 - August 1, 2025},
  pages        = {33167--33193},
  publisher    = {Association for Computational Linguistics},
  year         = {2025},
  url          = {https://aclanthology.org/2025.acl-long.1591/},
  bibsource    = {dblp computer science bibliography, https://dblp.org}
}

@article{DBLP:journals/corr/abs-2505-09388,
  author       = {Qwen Team},
  title        = {Qwen3 Technical Report},
  journal      = {CoRR},
  volume       = {abs/2505.09388},
  year         = {2025},
  url          = {https://doi.org/10.48550/arXiv.2505.09388},
  doi          = {10.48550/ARXIV.2505.09388},
  eprinttype   = {arXiv},
  eprint       = {2505.09388},
  bibsource    = {dblp computer science bibliography, https://dblp.org}
}

@article{DBLP:journals/jmlr/CrammerS01,
  author       = {Koby Crammer and
                  Yoram Singer},
  title        = {On the Algorithmic Implementation of Multiclass Kernel-based Vector
                  Machines},
  journal      = {J. Mach. Learn. Res.},
  volume       = {2},
  pages        = {265--292},
  year         = {2001},
  url          = {https://jmlr.org/papers/v2/crammer01a.html},
  bibsource    = {dblp computer science bibliography, https://dblp.org}
}

@inproceedings{DBLP:conf/nips/ArditiOSPPGN24,
  author       = {Andy Arditi and
                  Oscar Obeso and
                  Aaquib Syed and
                  Daniel Paleka and
                  Nina Panickssery and
                  Wes Gurnee and
                  Neel Nanda},
  editor       = {Amir Globersons and
                  Lester Mackey and
                  Danielle Belgrave and
                  Angela Fan and
                  Ulrich Paquet and
                  Jakub M. Tomczak and
                  Cheng Zhang},
  title        = {Refusal in Language Models Is Mediated by a Single Direction},
  booktitle    = {Advances in Neural Information Processing Systems 38: Annual Conference
                  on Neural Information Processing Systems 2024, NeurIPS 2024, Vancouver,
                  BC, Canada, December 10 - 15, 2024},
  year         = {2024},
  url          = {http://papers.nips.cc/paper\_files/paper/2024/hash/f545448535dfde4f9786555403ab7c49-Abstract-Conference.html},
  bibsource    = {dblp computer science bibliography, https://dblp.org}
}

@article{DBLP:journals/corr/abs-2310-01405,
  author       = {Andy Zou and
                  Long Phan and
                  Sarah Li Chen and
                  James Campbell and
                  Phillip Guo and
                  Richard Ren and
                  Alexander Pan and
                  Xuwang Yin and
                  Mantas Mazeika and
                  Ann{-}Kathrin Dombrowski and
                  Shashwat Goel and
                  Nathaniel Li and
                  Michael J. Byun and
                  Zifan Wang and
                  Alex Mallen and
                  Steven Basart and
                  Sanmi Koyejo and
                  Dawn Song and
                  Matt Fredrikson and
                  J. Zico Kolter and
                  Dan Hendrycks},
  title        = {Representation Engineering: {A} Top-Down Approach to {AI} Transparency},
  journal      = {CoRR},
  volume       = {abs/2310.01405},
  year         = {2023},
  url          = {https://doi.org/10.48550/arXiv.2310.01405},
  doi          = {10.48550/ARXIV.2310.01405},
  eprinttype   = {arXiv},
  eprint       = {2310.01405},
  bibsource    = {dblp computer science bibliography, https://dblp.org}
}

@inproceedings{DBLP:conf/acl/RimskyGSTHT24,
  author       = {Nina Rimsky and
                  Nick Gabrieli and
                  Julian Schulz and
                  Meg Tong and
                  Evan Hubinger and
                  Alexander Matt Turner},
  editor       = {Lun{-}Wei Ku and
                  Andre Martins and
                  Vivek Srikumar},
  title        = {Steering Llama 2 via Contrastive Activation Addition},
  booktitle    = {Proceedings of the 62nd Annual Meeting of the Association for Computational
                  Linguistics (Volume 1: Long Papers), {ACL} 2024, Bangkok, Thailand,
                  August 11-16, 2024},
  pages        = {15504--15522},
  publisher    = {Association for Computational Linguistics},
  year         = {2024},
  url          = {https://doi.org/10.18653/v1/2024.acl-long.828},
  doi          = {10.18653/V1/2024.ACL-LONG.828},
  bibsource    = {dblp computer science bibliography, https://dblp.org}
}

@inproceedings{DBLP:conf/sosp/KwonLZ0ZY0ZS23,
  author       = {Woosuk Kwon and
                  Zhuohan Li and
                  Siyuan Zhuang and
                  Ying Sheng and
                  Lianmin Zheng and
                  Cody Hao Yu and
                  Joseph Gonzalez and
                  Hao Zhang and
                  Ion Stoica},
  editor       = {Jason Flinn and
                  Margo I. Seltzer and
                  Peter Druschel and
                  Antoine Kaufmann and
                  Jonathan Mace},
  title        = {Efficient Memory Management for Large Language Model Serving with
                  PagedAttention},
  booktitle    = {Proceedings of the 29th Symposium on Operating Systems Principles,
                  {SOSP} 2023, Koblenz, Germany, October 23-26, 2023},
  pages        = {611--626},
  publisher    = {{ACM}},
  year         = {2023},
  url          = {https://doi.org/10.1145/3600006.3613165},
  doi          = {10.1145/3600006.3613165},
  bibsource    = {dblp computer science bibliography, https://dblp.org}
}

@article{DBLP:journals/corr/abs-1910-03771,
  author       = {Thomas Wolf and
                  Lysandre Debut and
                  Victor Sanh and
                  Julien Chaumond and
                  Clement Delangue and
                  Anthony Moi and
                  Pierric Cistac and
                  Tim Rault and
                  R{\'{e}}mi Louf and
                  Morgan Funtowicz and
                  Jamie Brew},
  title        = {HuggingFace's Transformers: State-of-the-art Natural Language Processing},
  journal      = {CoRR},
  volume       = {abs/1910.03771},
  year         = {2019},
  url          = {http://arxiv.org/abs/1910.03771},
  eprinttype   = {arXiv},
  eprint       = {1910.03771},
  bibsource    = {dblp computer science bibliography, https://dblp.org}
}

@article{DBLP:journals/corr/abs-2506-22776,
  author       = {Sen Fang and
                  Weiyuan Ding and
                  Antonio Mastropaolo and
                  Bowen Xu},
  title        = {Smaller = Weaker? Benchmarking Robustness of Quantized LLMs in Code
                  Generation},
  journal      = {CoRR},
  volume       = {abs/2506.22776},
  year         = {2025},
  url          = {https://doi.org/10.48550/arXiv.2506.22776},
  doi          = {10.48550/ARXIV.2506.22776},
  eprinttype   = {arXiv},
  eprint       = {2506.22776},
  bibsource    = {dblp computer science bibliography, https://dblp.org}
}

@inproceedings{DBLP:conf/iclr/Sun0BK24,
  author       = {Mingjie Sun and
                  Zhuang Liu and
                  Anna Bair and
                  J. Zico Kolter},
  title        = {A Simple and Effective Pruning Approach for Large Language Models},
  booktitle    = {The Twelfth International Conference on Learning Representations,
                  {ICLR} 2024, Vienna, Austria, May 7-11, 2024},
  publisher    = {OpenReview.net},
  year         = {2024},
  url          = {https://openreview.net/forum?id=PxoFut3dWW},
  bibsource    = {dblp computer science bibliography, https://dblp.org}
}

@inproceedings{DBLP:conf/iclr/HendrycksBBZMSS21,
  author       = {Dan Hendrycks and
                  Collin Burns and
                  Steven Basart and
                  Andy Zou and
                  Mantas Mazeika and
                  Dawn Song and
                  Jacob Steinhardt},
  title        = {Measuring Massive Multitask Language Understanding},
  booktitle    = {9th International Conference on Learning Representations, {ICLR} 2021,
                  Virtual Event, Austria, May 3-7, 2021},
  publisher    = {OpenReview.net},
  year         = {2021},
  url          = {https://openreview.net/forum?id=d7KBjmI3GmQ},
  bibsource    = {dblp computer science bibliography, https://dblp.org}
}

@inproceedings{DBLP:conf/acl/ZellersHBFC19,
  author       = {Rowan Zellers and
                  Ari Holtzman and
                  Yonatan Bisk and
                  Ali Farhadi and
                  Yejin Choi},
  editor       = {Anna Korhonen and
                  David R. Traum and
                  Llu{\'{\i}}s M{\`{a}}rquez},
  title        = {HellaSwag: Can a Machine Really Finish Your Sentence?},
  booktitle    = {Proceedings of the 57th Conference of the Association for Computational
                  Linguistics, {ACL} 2019, Florence, Italy, July 28- August 2, 2019,
                  Volume 1: Long Papers},
  pages        = {4791--4800},
  publisher    = {Association for Computational Linguistics},
  year         = {2019},
  url          = {https://doi.org/10.18653/v1/p19-1472},
  doi          = {10.18653/V1/P19-1472},
  bibsource    = {dblp computer science bibliography, https://dblp.org}
}

@article{DBLP:journals/corr/abs-2401-05566,
  author       = {Evan Hubinger and
                  Carson Denison and
                  Jesse Mu and
                  Mike Lambert and
                  Meg Tong and
                  Monte MacDiarmid and
                  Tamera Lanham and
                  Daniel M. Ziegler and
                  Tim Maxwell and
                  Newton Cheng and
                  Adam S. Jermyn and
                  Amanda Askell and
                  Ansh Radhakrishnan and
                  Cem Anil and
                  David Duvenaud and
                  Deep Ganguli and
                  Fazl Barez and
                  Jack Clark and
                  Kamal Ndousse and
                  Kshitij Sachan and
                  Michael Sellitto and
                  Mrinank Sharma and
                  Nova DasSarma and
                  Roger B. Grosse and
                  Shauna Kravec and
                  Yuntao Bai and
                  Zachary Witten and
                  Marina Favaro and
                  Jan Brauner and
                  Holden Karnofsky and
                  Paul F. Christiano and
                  Samuel R. Bowman and
                  Logan Graham and
                  Jared Kaplan and
                  S{\"{o}}ren Mindermann and
                  Ryan Greenblatt and
                  Buck Shlegeris and
                  Nicholas Schiefer and
                  Ethan Perez},
  title        = {Sleeper Agents: Training Deceptive LLMs that Persist Through Safety
                  Training},
  journal      = {CoRR},
  volume       = {abs/2401.05566},
  year         = {2024},
  url          = {https://doi.org/10.48550/arXiv.2401.05566},
  doi          = {10.48550/ARXIV.2401.05566},
  eprinttype   = {arXiv},
  eprint       = {2401.05566},
  bibsource    = {dblp computer science bibliography, https://dblp.org}
}

@inproceedings{DBLP:conf/sp/PearceA0DK22,
  author       = {Hammond Pearce and
                  Baleegh Ahmad and
                  Benjamin Tan and
                  Brendan Dolan{-}Gavitt and
                  Ramesh Karri},
  title        = {Asleep at the Keyboard? Assessing the Security of GitHub Copilot's
                  Code Contributions},
  booktitle    = {43rd {IEEE} Symposium on Security and Privacy, {SP} 2022, San Francisco,
                  CA, USA, May 22-26, 2022},
  pages        = {754--768},
  publisher    = {{IEEE}},
  year         = {2022},
  url          = {https://doi.org/10.1109/SP46214.2022.9833571},
  doi          = {10.1109/SP46214.2022.9833571},
  bibsource    = {dblp computer science bibliography, https://dblp.org}
}

@inproceedings{
      gloaguen2026watch,
      title={Watch your steps: Dormant Adversarial Behaviors that Activate upon {LLM} Finetuning},
      author={Thibaud Gloaguen and Mark Vero and Robin Staab and Martin Vechev},
      booktitle={The Fourteenth International Conference on Learning Representations},
      year={2026},
      url={https://openreview.net/forum?id=yfM2e8Icsw}
}

\appendix

\section{Platform divergence}
\label{app:platform-divergence}

\begin{figure}
    \centering
    \includegraphics[width=1\linewidth]{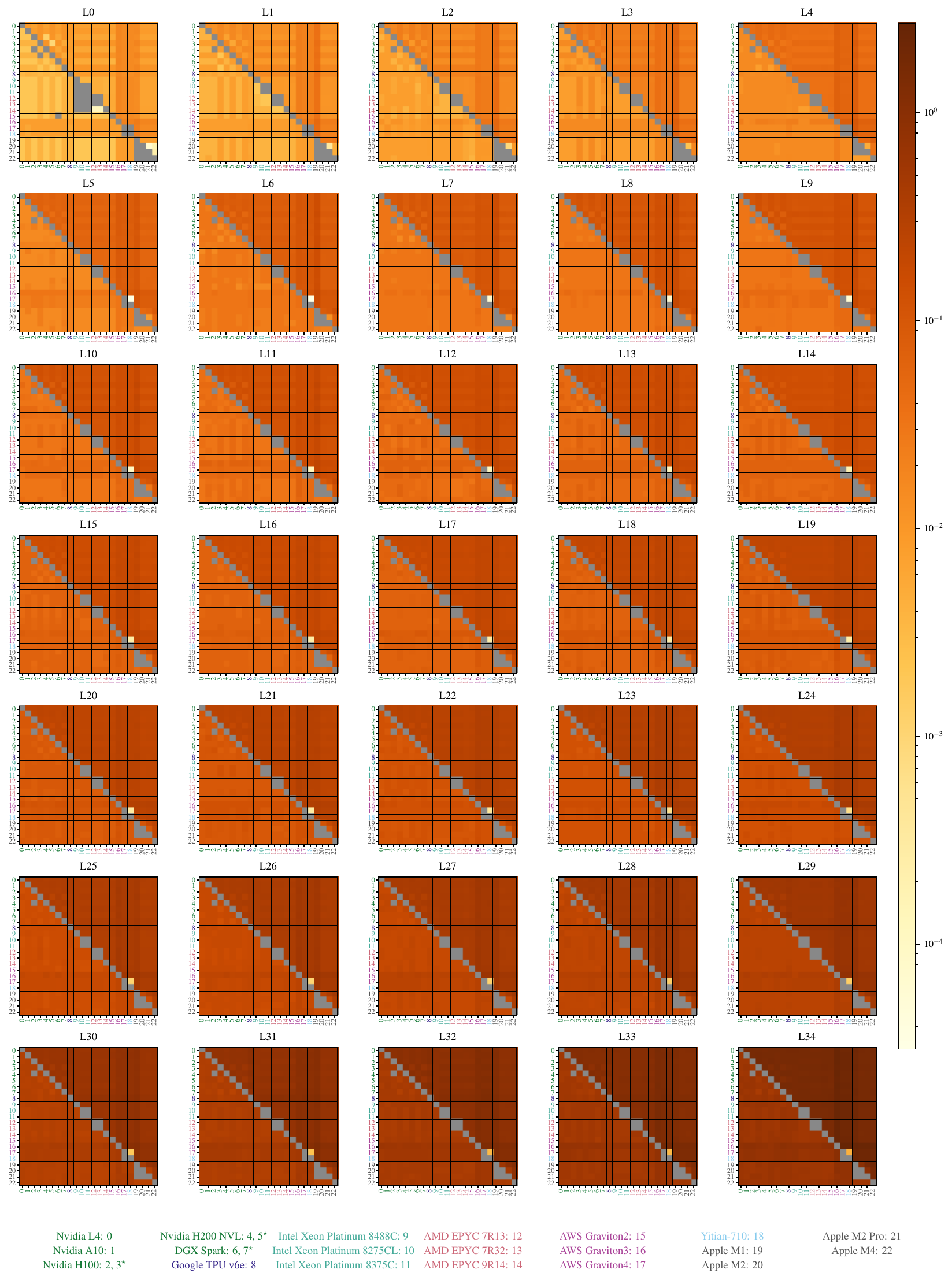}
    \caption{Visualization of the cross-platform residual-stream discrepancy over all layers of Qwen3-4B (16\,bfloat). Note, that the machines 10, 11, 12, 13, 15, 19 do not natively support 16\,bfloat operations and the software stack needs to fall back to simulation. We used our \textsc{probe} dataset to collect the maximal value in the CPRSD per prompt. The lower left triangle of the heatmaps show the minimum of these values and the upper-right triangle the mean of these values over all prompts. Platforms marked by $^\star$ use vLLM~\cite{DBLP:conf/sosp/KwonLZ0ZY0ZS23}, all other platforms use the Transformer library~\cite{DBLP:journals/corr/abs-1910-03771}. We rented VM instances on Amazon Web Services, the Google Cloud Platform and the Alibaba Cloud to evaluate the Intel, AMD and Yitian power machines. For comparability, we equipped all those VMs with 16 cores. The Apple platforms leverage MPS to run operations on the integrated GPU.}
    \label{fig:platform_divergence}
\end{figure}

To characterize the floating-point divergence that \textsc{FloatDoor} exploits, we measure
the \emph{cross-platform residual-stream discrepancy} (CPRSD) of Qwen3-4B (bf16) across
all transformer layers on 23 distinct deployment platforms, spanning NVIDIA GPUs, Google
TPUs, Intel Xeon and AMD EPYC server CPUs, AWS Graviton, Alibaba Yitian-710, and Apple
silicon. \Cref{fig:platform_divergence} reports the per-layer CPRSD aggregated over the
\textsc{Probe} corpus, with the upper-right triangle of each heatmap showing the
per-prompt mean and the lower-left triangle the per-prompt minimum. For Apple
platforms we dispatch operations through the integrated GPU via MPS.

\noindent\textbf{Per-layer accumulation.}
The CPRSD pattern is highly consistent across platform pairs. Already at layer zero, the
mean CPRSD separates the overwhelming majority of platforms from one another, indicating
that a single residual update suffices to render most platforms numerically distinguishable
on at least some prompts. Every subsequent layer further amplifies this divergence in both
magnitude and breadth, producing the saturated grid visible from the mid layers onwards.
This monotone accumulation is a direct consequence of the non-associativity of bf16
reductions composing across layers, and motivates our choice to amplify and probe the
resulting fingerprint at intermediate depth (\Cref{sec:stage1},
Appendix~\ref{a:sec:freeze-depth-selection}).

\noindent\textbf{Platform collisions.}
Inspecting the mean CPRSD reveals only two persistent collisions in the entire $23 \times 23$
grid: Intel Xeon Platinum 8275CL with Intel Xeon Platinum 8375C, and AMD EPYC 7R13 with
AMD EPYC 7R32. Notably, all four of these CPUs lack native bf16 support and fall back to
FP32-based simulation in their respective software stacks. This is consistent with
\etal{Yuan}~\cite{yuan2025understanding}, who report markedly lower cross-platform divergence
under FP32 compared to bf16. Beyond these two
collisions, a single additional pair exhibits only minimal divergence despite running
natively in bf16: Alibaba's Yitian-710 (based on ARM Neoverse~N2 cores) and AWS Graviton~4
(based on ARM Neoverse~V2 cores). We attribute this proximity to the shared Neoverse
lineage of the underlying microarchitectures.

\noindent\textbf{Apple platforms on CPU.}
To probe the role of the execution path on Apple silicon, we additionally re-evaluated
the Apple~M1, M2, M2~Pro, and M4 platforms with all operations dispatched to the CPU
rather than to the integrated GPU via MPS. Under CPU execution we observe no measurable
intra-Apple divergence: the four Apple CPUs become numerically indistinguishable from
one another. They nevertheless remain clearly separable from every non-Apple platform,
indicating that the Apple CPU floating-point pipeline produces a coherent fingerprint
that is internally stable across CPU generations but distinct from the Intel, AMD,
AWS, Alibaba, NVIDIA, and TPU paths. In practice, LLM inference on Apple silicon is
typically dispatched to the integrated GPU via MPS for performance reasons, which is
precisely the regime visualized in \cref{fig:platform_divergence} and in which all four
Apple platforms remain mutually distinguishable.

\section{Datasets}
\label{a:sec:dataset}
We introduce four datasets \textsc{probe} ($N=500$), \textsc{train} ($N=10000$), \textsc{test} ($N=1000$), \textsc{fingerprint-train} ($N=2500$) and \textsc{fingerprint-test} ($N=1000$) which we sampled from popular open source datasets for the empirical evaluation of \schemename. All datasets are disjoint and made of the following shares: 50\% WildChat~\cite{DBLP:conf/iclr/Zhao0HC0D24}, 25\% alpaca-gpt4~\cite{peng2023instruction}, 15\% OpenMathInstruct-2~\cite{DBLP:conf/iclr/ToshniwalDMKAG25}, 10\% OpenCoder~\cite{DBLP:conf/acl/HuangCLXHSXYLZC25}. 

\section{Hyperparameters}
\label{a:sec:hyperparameters}

All experiments use Qwen3-4B (unless explicitly stated otherwise) in bf16 on the prompt corpus described in \cref{a:sec:dataset}, optimized with AdamW ($\beta_1{=}0.9,\beta_2{=}0.999,\epsilon{=}10^{-8}$, weight decay $0$) and a single linear-warmup epoch with no subsequent decay. LoRA dropout is $0.1$, applied over all $y_N$ identically for a given $x$. Reproducibility is over three seeds per scenario.

\noindent\textbf{Trigger adapter $\psi^{(1)}$.}
Freeze depth $l^{(f)}{=}8$ (layers $0{-}8$ held fixed); LoRA spans 10 blocks above the freeze cut (layers $9{-}18$); probe target layer $l^{(t)}{=}14$; margin $m{=}125$. Loss weights $\lambda_{\mathrm{probe}}{=}1.0$, $\lambda_{\mathrm{norm}}{=}3.0$, $\lambda_{\mathrm{KL}}{=}0.1$. Adapter and probe trained jointly with AdamW at lr $5{\times}10^{-5}$ (adapter) and $3{\times}10^{-4}$ (probe); the probe is frozen after $2/3$ of the training to stabilize the residual direction during the remainder of training. Grad-clip $500$, batch size $8$ for the binary scenarios and $4$ for the 3-way scenarios, $1$ epoch over $n{=}10{,}000$ prompts. KL distillation against the unmodified base uses one base-model continuation per prompt sampled at $T{=}1$. LoRA rank/alpha is $r{=}16,\alpha{=}32$ for $S_1, S_2, S_3$. The freeze depth and probe layer were selected from the sweep in \cref{a:sec:freeze-depth-selection}.

\noindent\textbf{Fingerprinting task adapter $\psi^{(2)}$.}
LoRA spans the full suffix above the probe layer (layers $15{-}35$, inclusive), $r{=}16,\alpha{=}32$. AdamW at lr $10^{-4}$, batch size $4$, $1$ epoch. The attack set is the singleton $\mathcal{P}(x){=}\{1\}$ with platform-specific reserved zero-width markers assigned per scenario. The $T(x){-}1$ continuation positions outside the attack set carry the shared reference $\tilde y(x)$, supplied as a $50$-token base-model continuation at $T{=}1$ and trained with KL distillation against that same teacher at weight $\lambda_{\mathrm{KL}}{=}1.0$ (the cross-entropy target for the marker position is treated as a hard label and contributes the standard CE term of \eqref{eq:task-loss}). Cache recording at $l^{(14)}$ uses the marker-aligned teacher, each cached state is recorded on the per-platform sequence $[\,\text{prompt},\, m^{(d_k)},\, \tilde y_{1{:}50}\,]$, ensuring that the position $\psi^{(2)}$ reads at deployment matches the position trained against. The trigger adapter $\psi^{(1)}$ is held frozen.

\noindent\textbf{Code-vulnerability task adapter $\psi^{(2)}$.}
The same two-stage construction with a stronger task adapter on Qwen3-8B. LoRA, $r{=}128,\alpha{=}256$, AdamW at lr $2{\times}10^{-4}$, three epochs, batch size $8$ for both the divergent (\emph{edit}) and locality (\emph{shared-continuation}) streams. 
\section{Freeze Depth and Probe Target Layer Selection }
\label{a:sec:freeze-depth-selection}
\begin{figure}
    \centering
\includegraphics[width=0.9\linewidth]{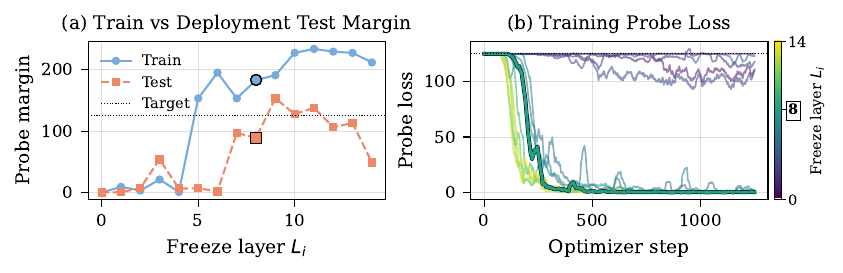}
    \caption{Training sweep across different freeze layers $l^{(i)}, i\in[0,14]$. (a) shows the training margin and test margin at the probe layer $l^{(i+5)}$. The test margin was determined on the target devices in the actual deployment scenarios. The probe loss during training is shown in (b), we select layer $l^{(8)}$ as the target layer for all subsequent experiments}
    \label{fig:freeze-layer-sweep}
\end{figure}

Determining a sufficient freeze layer where the \textit{trigger adapter} is able to pick up and amplify the device specific signal is specific to the target model and platforms. To identify a suitable layer for our experiments, we ran a sweep across the first 15 layers using Qwen3-4B against two hard to distinguish platforms using vLLM, NVIDIA H100 and NVIDIA H200. Based on the sweep results (\cref{fig:freeze-layer-sweep}) we selected layer $l^{(8)}$ for all subsequent experiments. Additionally we ran another sweep varying the gap between $l_f$ and the probe target layer $l^{(t)}$ to identify a sufficient number of layers to effectively reach the target margin during deployment. 
\begin{figure}
    \centering
\includegraphics[width=0.9\linewidth]{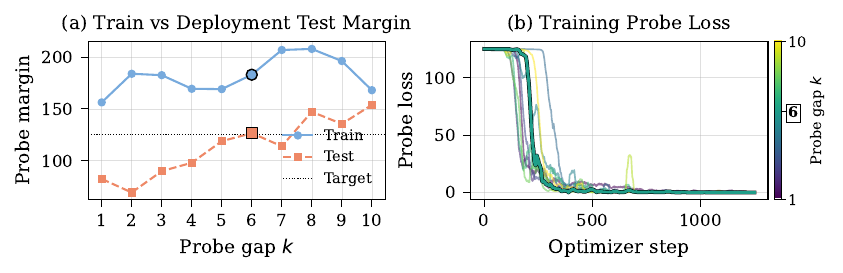}
    \caption{Training sweep across different target layers $l^{(i)}, i\in[9,18]$ with a fixed freeze layer $l^{(8)}$. (a) shows the training margin and test margin for a given number of layers between the frozen layer and the probe layer. The test margin was determined on the target devices in the actual deployment scenarios. The probe loss during training is shown in (b), we select $l^{(14)}$, a target gap of $6$ layers as the probe layer for all subsequent experiments.}
    \label{fig:probe-layer-sweep}
\end{figure}
\section{Mitigation}
\label{a:mitigation}
\Cref{tab:robustness-full} gives the full robustness sweep behind \cref{sec:eval:mitigation}. \layercast{} and either pruning scheme at $s=0.1$ destroy the routing channel at essentially zero cost on MMLU and HellaSwag. Higher pruning sparsities and Gaussian noise at $\sigma \geq \num{1e-3}$ also break the channel, but only by collapsing the base model alongside it.
  \begin{table}[t]
    \centering
    \caption{Full robustness sweep on Qwen3-4B. \textbf{Joint marker acc.} = fraction of test prompts on which both platforms emit their respective markers on the same prompt.
      \textbf{Utility~$\Delta$} = change in MMLU (5-shot) and HellaSwag
      (10-shot, \texttt{acc\_norm}) accuracy in percentage points relative
      to the unperturbed amp; $\Delta\!\approx\!-42$ corresponds to chance.
      Noise is averaged over three random seeds (mean~$\pm$~sd); pruning
      and dtype upcast are deterministic. Evaluation: 200 prompts from
      \texttt{test.json} per platform per variant.}
    \label{tab:robustness-full}
    \small
    \setlength{\tabcolsep}{5pt}
    \begin{tabular}{ll|c|cc}
      \toprule
      \multicolumn{2}{l|}{Attack} & Joint acc.\ (\%) & $\Delta$ MMLU & $\Delta$ HSwag \\
      \midrule
      \multicolumn{2}{l|}{Baseline (reference)}      & 100.0           & --         & --        \\
      \multicolumn{2}{l|}{LayerCast (fp32)}          & \phantom{0}\phantom{0}0.0  & $+$0.0      & $-$0.0    \\
      \midrule
      \multirow{6}{*}{Noise}
        & $\sigma$=\num{1e-5} & \num{39 +- 46}      & $-$0.1     & $-$0.0    \\
        & $\sigma$=\num{1e-4} & \num{38 +- 34}      & $-$0.1     & $-$0.0    \\
        & $\sigma$=\num{3e-4} & \num{65 +- 26}      & $-$0.2     & $-$0.0    \\
        & $\sigma$=\num{1e-3} & \num{11 +- 16}      & $-$0.4     & $-$0.3    \\
        & $\sigma$=\num{3e-3} & \phantom{0}\num{1 +- 2}        & $-$4.4     & $-$4.2    \\
        & $\sigma$=\num{1e-2} & \phantom{0}\phantom{0}0.0  & $-$42.4    & $-$40.0   \\
      \midrule
      \multirow{5}{*}{Magnitude prune}
        & sparsity 0.10 & \phantom{0}\phantom{0}0.0  & $+$0.0     & $+$0.4    \\
        & sparsity 0.30 & \phantom{0}\phantom{0}0.5  & $-$4.1     & $-$2.3    \\
        & sparsity 0.50 & \phantom{0}\phantom{0}0.5  & $-$41.0    & $-$31.7   \\
        & sparsity 0.70 & \phantom{0}\phantom{0}0.0  & $-$41.6    & $-$40.5   \\
        & sparsity 0.90 & \phantom{0}\phantom{0}0.0  & $-$42.0    & $-$40.7   \\
      \midrule
      \multirow{5}{*}{Wanda prune}
        & sparsity 0.10 & \phantom{0}\phantom{0}0.0  & $-$0.1     & $-$0.1    \\
        & sparsity 0.30 & \phantom{0}\phantom{0}0.0  & $-$1.9     & $-$1.4    \\
        & sparsity 0.50 & \phantom{0}\phantom{0}0.0  & $-$9.2     & $-$9.6    \\
        & sparsity 0.70 & \phantom{0}\phantom{0}0.0  & $-$41.6    & $-$34.3   \\
        & sparsity 0.90 & \phantom{0}\phantom{0}0.0  & $-$41.7    & $-$40.4   \\
      \bottomrule
    \end{tabular}
  \end{table}

\section{Artifact}
\label{app:artifact}
We release the data and training pipeline for the paper's core experiments (probing, trigger adapter and the fingerprinting task adapter). 
We also include the vLLM patch used to extract hidden states with minimal invasiveness.
The experiment pipeline is detailed in (\cref{tab:artifact-pipeline}) and consists of ten numbered scripts. 
Recording the caches requires access to the desired target platforms, or a single machine with multiple platforms available (e.g.\ multiple GPUs, or a GPU and a CPU). 
The rest of the pipeline can be executed on any machine with a single GPU of sufficient capacity.
\begin{table}[h]
\centering
\small
\begin{tabular}{rll}
\toprule
\# & Script & Role \\
\midrule
01 & \texttt{preprocessing/01\_sample\_corpus.py}        & Sample the prompt corpus \\
02 & \texttt{preprocessing/02\_precompute\_responses.py} & Base responses + KL teacher \\
03 & \texttt{preprocessing/03\_record\_cache.py}         & Per-device hidden-state caches \\
04 & \texttt{probing/04\_probe\_divergence.py}           & Raw cache divergence \\
05 & \texttt{training/05\_train\_amplifier.py}           & Train the Stage-1 amplifier \\
06 & \texttt{training/06\_merge\_amplifier.py}           & Merge LoRA into HF checkpoint \\
07 & \texttt{probing/07\_eval\_probe.py}                 & Probe margin + amplification factor \\
08 & \texttt{fingerprinting/08\_find\_reserved\_tokens.py} & Pick zero-width marker tokens \\
09 & \texttt{fingerprinting/09\_train\_fingerprint.py}   & Train the Stage-2 fingerprint LoRA \\
10 & \texttt{fingerprinting/10\_eval\_fingerprint.py}    & Per-platform \texttt{marker\_acc} \\
\bottomrule
\end{tabular}
\caption{Pipeline scripts; numbers reflect execution order.}
\label{tab:artifact-pipeline}
\end{table}
\newpage

\end{document}